\RequirePackage{fix-cm}
\documentclass[twocolumn,epjc3]{svjour3}
\RequirePackage{graphicx,subfigure}

\begin{document}
\title{First test of an enriched $^{116}$CdWO$_4$ scintillating bolometer for neutrinoless double-beta-decay searches}
\titlerunning{enrCWO bolometer}

\author{A.S.~Barabash\thanksref{addr1}\and
F.A.~Danevich\thanksref{addr2}\and
Y.~Gimbal-Zofka\thanksref{addr3}\and
A.~Giuliani\thanksref{addr3,addr4,e1}\and 
S.I.~Konovalov\thanksref{addr1}\and
M.~Mancuso\thanksref{addr3,addr4,e2}\and
P.~de~Marcillac\thanksref{addr3}\and
S.~Marnieros\thanksref{addr3}\and C.~Nones\thanksref{addr5}\and
V.~Novati\thanksref{addr3}\and E.~Olivieri\thanksref{addr3}\and
G. Pessina\thanksref{addr6} \and
D.V.~Poda\thanksref{addr2,addr3} \and
O.G.~Polischuk\thanksref{addr2} \and
V.N.~Shlegel\thanksref{addr7}\and
V.I.~Tretyak\thanksref{addr2,addr8}\and
V.I.~Umatov\thanksref{addr1}\and
A.S.~Zolotarova\thanksref{addr5}}
\thankstext{e1}{e-mail: andrea.giuliani@csnsm.in2p3.fr}
\thankstext{e2}{Presently at Max-Planck-Institut f\"ur Physik, 80805 Munich, Germany}
\institute{National Research Centre ``Kurchatov Institute'', ITEP, 117218 Moscow, Russia \label{addr1}
\and
Institute for Nuclear Research, MSP 03680 Kyiv, Ukraine \label{addr2}
\and
CSNSM, Univ. Paris-Sud, CNRS/IN2P3, Universit\'e Paris-Saclay, 91405 Orsay, France \label{addr3}
\and
DISAT, Universit\`a dell'Insubria, 22100 Como, Italy \label{addr4}
\and
CEA Saclay, DSM/IRFU, 91191 Gif-sur-Yvette Cedex, France \label{addr5}
\and
INFN, Sezione di Milano Bicocca, 20126 Milano, Italy \label{addr6}
\and
Nikolaev Institute of Inorganic Chemistry, 630090 Novosibirsk, Russia \label{addr7}
\and
INFN, sezione di Roma, I-00185 Rome, Italy \label{addr8}}

\maketitle

\begin{abstract}
For the first time, a cadmium tungstate crystal scintillator enriched in $^{116}$Cd has been succesfully tested as a scintillating bolometer. The measurement was performed above ground at a temperature of 18 mK. The crystal mass was 34.5~g and the enrichment level $\sim 82$\%. Despite a substantial
pile-up effect due to above-ground operation, the detector demonstrated a high energy resolution ($2-7$~keV FWHM in $0.2-2.6$~MeV $\gamma$ energy range), a powerful particle identification capability and a high level of internal
radiopurity. These results prove that cadmium tungstate is an
extremely promising detector material for a next-generation neutrinoless
double-beta decay bolometric experiment, like that proposed in the
CUPID project (CUORE Upgrade with Particle IDentification).
\end{abstract}



\section{Introduction}

Neutrinoless double-beta ($0\nu2\beta$) decay is a hypothetical
nuclear transformation that changes the lepton number by two
units when a candidate even-even nucleus emits two electrons with no neutrino in the final state. The observation of $0\nu2\beta$ decay would testify lepton number non-conservation and the presence of a Majorana term in neutrino masses, and give information on the neutrino-mass absolute scale along with the ordering of the neutrino-mass eigenstates \cite{Vergados:2012,Barea:2012,Rodejohann:2012}. It
should be stressed that many effects beyond the Standard Model can contribute to the $0\nu2\beta$
decay rate \cite{Deppisch:2012,Pas:2015,Bilenky:2015,DellOro:2016}.

In contrast with the two-neutrino mode ($2\nu2\beta$),
experimentally observed in eleven isotopes with half-lives in
the range 10$^{18}$--10$^{24}$ yr (see reviews
\cite{Saakyan:2013,Barabash:2015,Barabash:2015a} and references
therein) and allowed in the Standard Model, the $0\nu2\beta$ decay has not been detected yet. The most
sensitive experiments give only half-life limits on the level
of $T_{1/2} > 10^{25}-10^{26}$ yr, which correspond to constraints on the
effective Majorana neutrino mass around $\langle m_{\nu}\rangle < 0.1-1$ eV, in the degenerate hierarchy region of the
neutrino mass eigenstates (see reviews
\cite{DellOro:2016,Barabash:2015a,Giuliani:2012,Cremonesi:2014,Sarazin:2015} and
the recent KamLAND-Zen result \cite{Gando:2016}). The goal of the
next-generation $0\nu2\beta$ experiments is to probe the inverted
hierarchy region of the neutrino mass ($\langle m_{\nu}\rangle
\sim 0.05-0.02$~eV). This neutrino mass scale corresponds to
half-lives $T_{1/2}\sim 10^{27}-10^{28}$ yr even for the nuclei
with the highest decay probability
\cite{Vergados:2012,Barea:2012}. The attainment of a so high sensitivity
requires the construction of a detector containing a large number of
$2\beta$ active nuclei ($10^3-10^4$ moles of isotope of interest),
extremely low (ideally zero) radioactive background, high
detection efficiency (obtainable in the calorimetric approach
``source = detector'') and ability to distinguish the effect
searched for (in particular, as high as possible energy
resolution). Taking into account the extremely low decay
probability and the difficulties of the nuclear matrix elements
calculations \cite{Vergados:2012,Barea:2012}, the experimental
program should include a few candidate nuclei.

The technique of low temperature scintillating bolometers looks
very promising to satisfy the above mentioned requirements
\cite{Pirro:2006,Beeman:2012,Artusa:2014}. The nucleus $^{116}$Cd
is one of the most attractive candidates thanks to one of the
highest energy release ($Q_{2\beta}$ = 2813.50(13)~keV
\cite{Rahaman:2011}), comparatively large natural isotopic
abundance ($\delta$ = 7.512(54)\% \cite{Meija:2016}),
applicability of centrifugation for cadmium isotopes enrichment in
a large amount, and availability of cadmium tungstate crystal
scintillators (CdWO$_4$).

Cadmium tungstate crystals are routinely produced on an industrial basis and are among the most radiopure and efficient scintillators, with a long history of applications in low counting experiments to search for double-beta decay
\cite{Danevich:1989,Danevich:1995,Danevich:1996a,Danevich:2003a,Belli:2008}
and investigate rare $\alpha$ \cite{Danevich:2003b} and $\beta$ decays
\cite{Alessandrello:1993,Danevich:1996b,Belli:2007}.
Recently, high-quality radiopure CdWO$_4$ crystal scintillators
were developed from deeply-purified cadmium samples enriched
in the isotopes $^{106}$Cd \cite{Belli:2010} and $^{116}$Cd
\cite{Barabash:2011} with the help of the low-thermal-gradient
Czochralski crystal-growth technique \cite{Grigoriev:2014}. These
enriched scintillators are currently and succesfully used in the $0\nu2\beta$ decay 
experiments with $^{106}$Cd \cite{Belli:2012,Belli:2016} and
$^{116}$Cd \cite{Poda:2014,Danevich:2016}. Important advantages
of the low-thermal-gradient Czochralski method are a high yield of
the crystal boules ($\approx87\%$) and an acceptable low level of
irrecoverable losses of enriched cadmium ($\approx2\%$). Thus,
production of high quality radiopure cadmium tungstate crystal
scintillators from enriched isotopes is already a well developed technique.
Starting from the beginning of nineties of the last century, CdWO$_4$ was intensively tested first as a pure bolometer \cite{Alessandrello:1993} and then as a scintillating bolometer with a high performance in terms of energy resolution, particle discrimination ability and low radioactive background
\cite{Pirro:2006,Gorla:2008,Gironi:2009,Arnaboldi:2010}.

The aforementioned results played a crucial role in including
CdWO$_4$ in the list of the possible candidates for the CUPID project
\cite{CUPID}. In this context, the first bolometric test of an
enriched $^{116}$CdWO$_4$ scintillating bolometer -- here reported -- adds a crucial missing piece of information in view of the full
implementation of the cadmium tungstate technology for $0\nu2\beta$ search. It should be stressed that reproducing the results achieved with materials of
natural isotopic composition with enriched crystal scintillators
is not trivial. Indeed, the procedures of purification of enriched
isotopes and the growth of crystals from enriched materials are
severely constrained by the strong requirements of a high yield in developing ready-to-use crystals and minimal losses of the costly enriched materials.
These requirements may affect negatively the bolometric
performance and the intrinsic background, which need to be specifically studied for bolometers containing enriched isotopes. Among the three candidates that are very attractive for the scintillating
bolometer technology, i.e. $^{100}$Mo, $^{82}$Se and $^{116}$Cd,
positive tests on enriched materials were performed before this
work only in the first two cases~\cite{Barabash:2014,Artusa:2016}.
The results here described on $^{116}$Cd complete the
investigation of these isotopes and enhance the unrivaled merits of the $^{116}$CdWO$_4$ technology.

\section{Test of a $^{116}$CdWO$_4$ scintillating bolometer}

\begin{figure}[htb]
\centering
\includegraphics[width=0.48\textwidth]{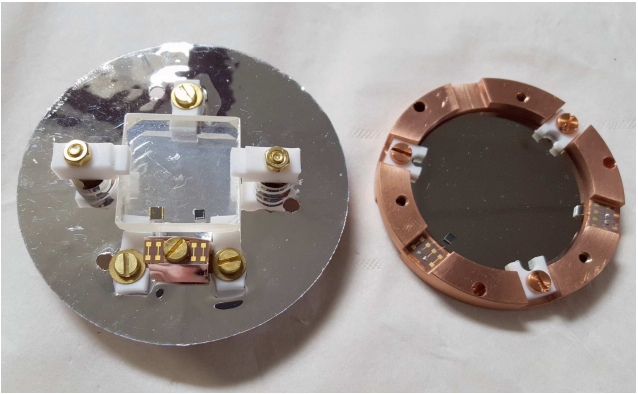}
\caption{Photograph of the 34.5~g $^{116}$CdWO$_4$ scintillating bolometer
assembled on a copper plate covered by a reflecting foil (left) together with
the Ge-based light detector (right). See the text for the details.}
\label{fig:detector}
\end{figure}

A sample of enriched $^{116}$CdWO$_4$ crystal scintillator
was cut from the wide part of the growth cone
of a 1.9~kg crystal boule~\cite{Barabash:2011} (see Fig.1 in
Ref.~\cite{Barabash:2016}, where the boule and cut parts are
shown). The crystal mass and size are respectively 34.5~g and $28\times27\times 6$ mm, and the isotopic concentration of $^{116}$Cd is 82\%. The light detector (LD) consists of a high-purity germanium wafer ($\oslash$44$\times$0.175 mm) produced by Umicore. The scintillator and the Ge wafer were fixed in individual copper frames by using PTFE pieces and brass / copper screws. The inner
surface of the detector holder was covered by a reflecting foil
(Vikuiti$^{\rm TM}$ Enhanced Specular Reflector Film) to improve the scintillation light
collection. A neutron transmutation doped (NTD) Ge thermistor with
a mass of $\sim 50$~mg was glued on the $^{116}$CdWO$_4$ crystal
by six spots of epoxy (Araldite\textregistered) to register the
temperature pulses induced by the absorption of particles in the
$^{116}$CdWO$_4$ crystal. An approximately three-times-smaller NTD
Ge thermistor was attached to the LD with the aim to reduce the
added heat capacity and to increase the LD sensitivity. Both
bolometers were supplied with a silicon chip on top of which a heavily doped
meander was formed by donor ion implantation. The meander resistance is
stable down to millikelvin temperature and was used as a heater
\cite{Andreotti:2012} to inject periodically fixed amounts of
thermal energy for the detector stabilization. The partially
assembled $^{116}$CdWO$_4$ scintillating bolometer and the LD are
shown in Fig. \ref{fig:detector}.

The low-temperature tests of the $^{116}$CdWO$_4$ scintillating
bolometer were performed in a cryogenic laboratory of the CSNSM
(Orsay, France) by using a dry high-power dilution
refrigerator~\cite{Mancuso:2014} with a 4~K stage cooled by a
pulse-tube. The sample holder is mechanically decoupled from the
mixing chamber by four springs to reduce the acoustic noise caused
by the cryostat vibrations. The outer vacuum chamber of the
refrigerator is surrounded by a passive shield made of low
radioactivity lead (10~cm minimum thickness) to suppress the
environmental $\gamma$ background. The shield mitigates the
pile-up problem typical for above-ground measurements with
macro-bolometers, given the slow response of these devices (tens
or even hundreds of milliseconds). For the same reason, we have used a
relatively small $^{116}$CdWO$_4$ sample aiming to reduce the counting rate of the environmental
$\gamma$ background.

A low-noise electronics based on DC-coupled voltage-sensitive amplifiers~\cite{Arnaboldi:2002} and located inside a Faraday
cage was used in the experiment. The $^{116}$CdWO$_4$ and the LD
NTD sensors were biased with currents of 4.2~nA and 25~nA,
respectively. The bias current was injected through two load
resistors in series with a total resistance of 200~M$\Omega$ for
both channels.  The stream data were filtered by a Bessel filter
with a high frequency cut-off at 675~Hz and acquired by a 16~bit
ADC with 10~kHz sampling frequency.

Most of the measurements were performed with the sample holder
temperature stabilized at 18.0 mK. However, the $^{116}$CdWO$_4$
detector was approximately 2~mK warmer due to a not reached
temperature equilibrium between the mixing chamber and the
detector itself, because the scintillating bolometer was mounted to
the mechanically-decoupled holder by means of brass rods, non-optimal for thermalization. Therefore, the NTD-Ge-thermistor
resistances ($R_{NTD}$) at the working temperature had a clear
trend to increase. For instance, the resistance of the heat
channel thermistor changed from an initial $\sim 0.4$~M$\Omega$ value to a final $\sim 1$~M$\Omega$ during the two-week background run. It is worth noting
that the sample-holder temperature reached 9.6~mK during a short
test with unregulated temperature, and the corresponding NTD-Ge-thermistor resistance of the $^{116}$CdWO$_4$ bolometer went
quickly up to 1.6~M$\Omega$ with a tendency to further increase.
We expect that a better thermal coupling and operation at lower
temperatures would enable much higher detector performance (see
the next Section). In this regard, we remark that the CUPID experiment
is expected to be performed at $\sim$10~mK base temperature, which
is in fact the value used in Cuoricino and CUORE-0, predecessors
of the CUORE experiment.

We accumulated 59.6 h data with a $^{232}$Th source (consisting of
a 15.2 g thoriated tungsten rod containing 1\% of Th), and 190.1 h
of background-only measurements, which altogether constitute 249.7
h life time. The $^{116}$CdWO$_4$ detector was calibrated by means
of the $\gamma$ quanta from the environmental radioactivity
(mainly emitted by $^{214}$Pb and $^{214}$Bi radionuclides from
the $^{238}$U chain) and in calibration run by $\gamma$ quanta
from the $^{232}$Th source (mainly $^{228}$Ac and $^{208}$Tl,
daughters of $^{232}$Th). The rear side of the LD was permanently
irradiated by a weak $^{55}$Fe X-ray source. In addition, an optic
fiber was mounted inside the cryostat to transmit LED light
pulses to the LD every 30~s, which can be also used for
calibration / stabilization purposes.

\section{Results and discussion}

The collected data were processed off-line by applying the optimum
filtering procedure \cite{Gatti:1986} and several
pulse-characterizing parameters were evaluated for each recorded
signal: the pulse amplitude, the rise- ($\tau_R$) and decay-
($\tau_D$) times\footnote{Here the rise-time is defined as the
time interval between 10\% an 90\% of the maximum amplitude of the
signal for the rising edge, while the decay-time corresponds to
the time interval between 90\% an 30\% of the maximum amplitude of
the signal for the decaying edge.}, several pulse-shape
indicators, and the DC baseline level of the pre-triggered part (over
0.15 s). In addition, the energy resolution of the filtered
baseline noise (FWHM$_{Bsl}$) and the amplitude of the signal
($S_{NTD}$) for a given deposited energy were estimated for each
data set ($1-3$ days of measurements). Some of these parameters,
characterizing the performance of the $^{116}$CdWO$_4$
scintillating bolometer and the LD, are given in Table
\ref{tab:performance}.

\begin{table}[!htb]
\caption{Technical data (see the text) for the $^{116}$CdWO$_4$ scintillating bolometer tested above ground at 18.0 mK (stabilized temperature of the sample holder). The $R_{NTD}$ and $S_{NTD}$ parameters correspond to the coldest conditions of the detector obtained at the end of the measurements. $\gamma$($\beta$) events registered by the $^{116}$CdWO$_4$ bolometer in the energy range $0.6-2.7$~MeV and the corresponding scintillation light signals detected by the LD in the energy range $\sim 15-85$~keV were used to evaluate the $\tau_R$ and $\tau_D$ parameters.} \footnotesize \centering
\begin{tabular}{cccccc}
\hline
Detector                 & $R_{NTD}$   & $S_{NTD}$ & FWHM$_{Bsl}$ & $\tau_R$ & $\tau_D$ \\
~                        & M$\Omega$   & nV/keV    & keV          & ms       & ms \\
\hline
LD                       & 0.12        & 258       & 0.6          &  1.3     & 4.7 \\
$^{116}$CdWO$_4$ & 1.0         & 135       & 1.5          &  5.1     & 28.5 \\
\hline
\end{tabular}
\label{tab:performance}
\end{table}

Taking into account the expected high light yield\footnote{Here we
define ``light yield'' the ratio between light and heat signal
amplitudes (converted into detected energy), which of course is lower than the absolute light yield of CdWO$_4$.} of cadmium tungstate at low temperatures
(e.g., $\sim$17 keV/MeV \cite{Arnaboldi:2010}), we have chosen a light
detector with a relatively modest performance, as it is visible
from Table \ref{tab:performance}. Therefore, we were not able to
separate clearly the $^{55}$Fe X-ray doublet (at 5.9 and 6.5 keV)
from the noise due to the poor energy resolution (FWHM$_{Fe55}$
$\approx 0.7$ keV). However, the LD time characteristics ($\tau_R$
and $\tau_D$ of the scintillation signals) are similar to that of
devices instrumented with small-size NTD Ge sensors (e.g., see the
performance of a first batch of six LDs preliminary tested for the
CUPID-0 detector array with Zn$^{82}$Se scintillating bolometers
\cite{Artusa:2016}).

\begin{figure}
\centering
\includegraphics[width=0.48\textwidth]{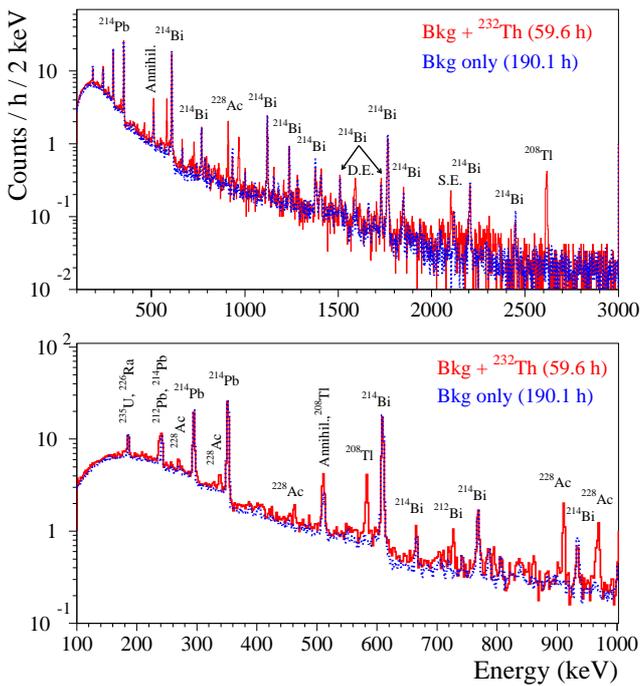}
\caption{In the top panel, the energy spectra of $\gamma$($\beta$) events
accumulated by the 34.5 g $^{116}$CdWO$_4$ bolometer in a $\sim 190$~h background run
(blue dotted histogram, Bkg only) and in a $\sim 60$~h calibration run (red solid histogram,
Bkg + $^{232}$Th) performed above ground at CSNSM. The nuclides originating the observed peaks are specified. ``D.E.'' and ``S.E.'' labels refer to double-escape and single-escape peaks related to the 2615~keV full-energy $\gamma$ peak of $^{208}$Tl. In the bottom panel, details of the spectra in the $100-1000$~keV energy range.}
\label{fig:spectrum}
\end{figure}

The performance of the $^{116}$CdWO$_4$ bolometer during the tests
is characterised by a high sensitivity $S_{NTD}$ and a quite low baseline noise (see Table \ref{tab:performance}). The heat-pulse profile of the $^{116}$CdWO$_4$ detector, as well as the sensitivity $S_{NTD}$, are similar to those observed in low-temperature tests with CdWO$_4$ bolometers produced from cadmium with natural isotopic composition \cite{Pirro:2006,Alessandrello:1993,Gorla:2008,Gironi:2009,Arnaboldi:2010}. This confirms that CdWO$_4$ is an excellent bolometric material and that the detector performance is not spoiled by the Cd isotopical enrichement. The energy spectra acquired with the $^{116}$CdWO$_4$ bolometer in background ($\sim 190$~h) and calibration ($\sim 60$~h) runs, shown in Fig.~\ref{fig:spectrum}, contain a number of sharp $\gamma$ peaks; even small-intensity (a few \%) $\gamma$ quanta of $^{214}$Bi are well visible, which altogether demonstrate an excellent spectrometric performance of the detector. The energy resolution (FWHM) varies from 2.9(1) keV at
242.0 keV ($\gamma$ quanta of $^{214}$Pb) to 8.3(9) keV at 2614.5 keV ($\gamma$ quanta of $^{208}$Tl).

\begin{figure}[htb]
\centering
\includegraphics[width=0.48\textwidth]{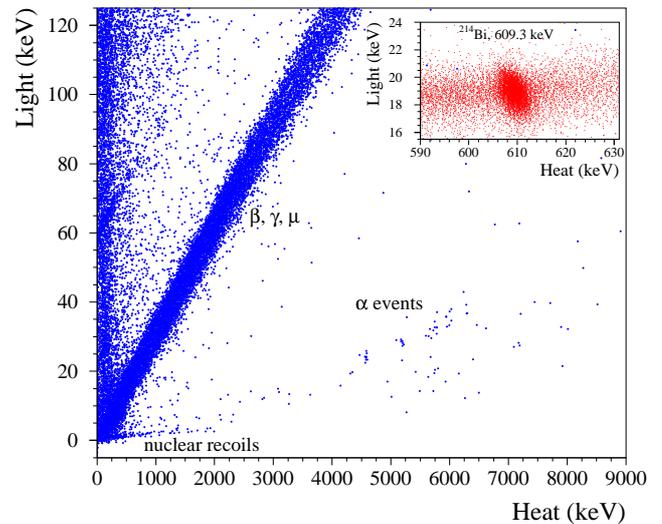}
\caption{Scatter plot of the light-versus-heat signals collected in a $\sim 250$~h
run with the 34.5~g $^{116}$CdWO$_4$ scintillating
bolometer. (Inset) The low energy part of the scatter plot in the
proximity of the 609.3 keV $\gamma$ peak of $^{214}$Bi. Light-heat
anticorrelation is clearly visible as a negative slop of the
609.3 keV cluster.} \label{fig:scatter}
\end{figure}

\begin{figure}[htb]
\centering
\includegraphics[width=0.48\textwidth]{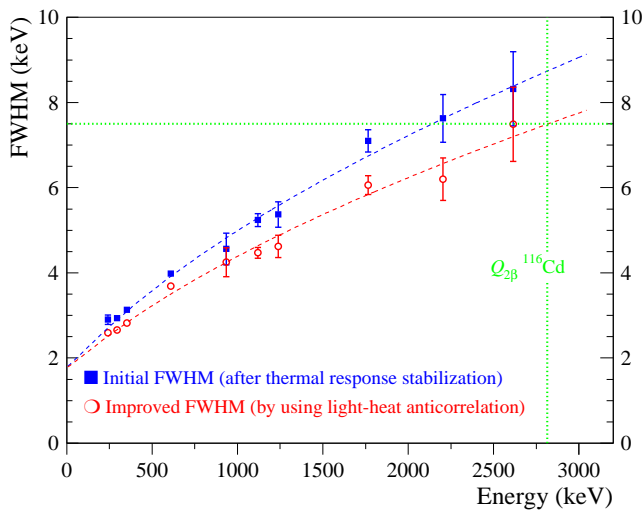}
\caption{Energy resolution (FWHM) of the $^{116}$CdWO$_4$ bolometric detector
after the stabilization of its thermal response by using a heater (blue filled
rectangles). The energy resolution improves by considering light-heat anticorrelation (red open circles). The fits of the data by a function FWHM = $\sqrt{a^2 + (b \times E_{\gamma})^c}$ (where FWHM and energy $E_{\gamma}$ are in keV; $a$, $b$, and $c$ are free parameters) are shown by dashed lines. The dotted lines indicate FWHM = 7.5 keV expected at $Q_{2\beta}$ of $^{116}$Cd.}
\label{fig:fwhm}
\end{figure}

The LD data were also processed with the trigger records of the
$^{116}$CdWO$_4$ data with an adjusted time difference between
the two channels (due to the longer rise-time of the
$^{116}$CdWO$_4$ heat signals) to search for coincidences. A
scatter plot of the pulse amplitudes of coincident heat and light
signals is shown in Fig.~\ref{fig:scatter}. The structures
visible on this figure are associated with $\gamma$($\beta$) and
cosmic muons interactions in the scintillator, bulk or/and surface
trace contamination by $\alpha$ radioactive nuclides from U/Th
chains, nuclear recoils due to ambient neutron scattering on the
nuclei in the $^{116}$CdWO$_4$ crystal, events with a prevailing
interaction in the light detector, or/and pile-up events. An
event-by-event analysis of the population distributed just below
the clusters in the $\alpha$ band demonstrates that these sporadic events
are affected by a signal overlapping, which can produce a
single-like event in the heat channel but a clear pile-up in the
light channel because of the much shorter time response of the
latter. The data exhibit anticorrelation between light and heat
signal amplitudes, as illustrated in the inset of Fig.
\ref{fig:scatter}. This feature was already observed in CdWO$_4$
scintillating bolometers based on cadmium with natural isotopic composition and can be used to enhance the energy resolution of the heat channel \cite{Arnaboldi:2010}. The improvement
is shown in Fig. \ref{fig:fwhm}, where the FWHM values of the most
intensive $\gamma$ peaks before and after applying the
anticorrelation correction are presented. It is evident from Fig.
\ref{fig:fwhm} that the achieved improvement is quite modest
(around 10\%) in contrast to the results of
Refs.~\cite{Gorla:2008,Arnaboldi:2010}. This may be explained by
a higher uniformity of the light collection from our smaller sample,
which is expected to make the light-heat anticorrelation less
significant. The energy resolution can be improved
further in an underground cryostat shielded against environmental
$\gamma$ radiation.

\begin{figure}[htb]
\centering
\includegraphics[width=0.48\textwidth]{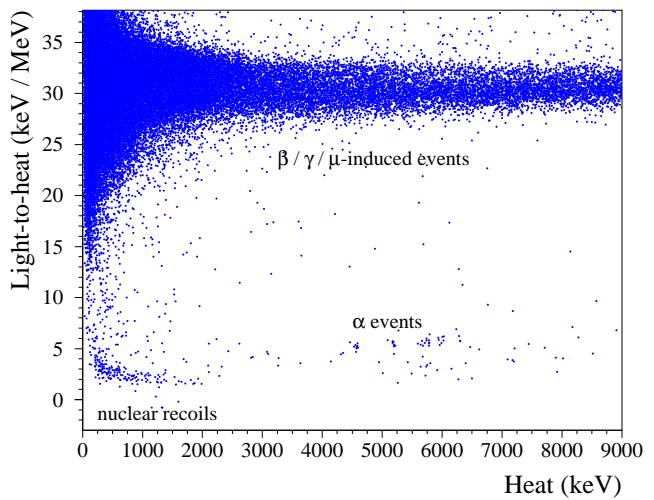}
\caption{The energy dependence of the light yield measured over
$\sim 250$~h with the 34.5 g $^{116}$CdWO$_4$ scintillating bolometer.}
\label{fig:qplot}
\end{figure}

The data of the heat-light coincidences can be transformed into
the so-called {\sl Q-plot} shown in Fig.~\ref{fig:qplot}. The
projection of the points on the y-axis can be used to evaluate
the light yield (LY) for different classes of registered events.
The LY for $\gamma$ quanta, $\beta$ particles and cosmic muons in
the energy interval $0.6-2.7$~MeV is $\sim 31$ keV/MeV; the LY for alpha particles with energy $4-7$~MeV (with the energy scale determined by a $\gamma$ calibration) is 5.5(1) keV/MeV, while the LY for nuclear recoils is even less, i.e. 2.6(1) keV/MeV, because of the further quenching of the scintillation light for heavier ions (a comprehensive study
of this phenomenon for cadmium tungstate can be found in
Refs.~\cite{Bizzeti:2012,Tretyak:2010}). It is worth to note that
such high LY values have never been reported for CdWO$_4$-based
scintillating bolometers. In particular, approximately twice lower
values were obtained in Ref.~\cite{Arnaboldi:2010} (however, the
crystal used in that study was an order of magnitude larger
in volume). This excellent result is obtained by the twice-larger
area of the LD, the overall compact geometry of the arrangement
(which enhances the light collection), a high optical transmittance of the
material~\cite{Barabash:2011}, and a low self-absorption of the scintillation photons in our
relatively thin sample. By using the LY's, one can also estimate
the quenching factors for $\alpha$'s and nuclear recoils as
0.175(3) and 0.084(3), respectively.

To evaluate the discrimination power (DP) between $\gamma(\beta)$
and $\alpha$ event distributions, the LY data shown in Fig.
\ref{fig:qplot} were used within the $2.6-7$~MeV energy range and
the $4-38$~keV/MeV LY interval (cutting most of the pile-up events in
the vicinity of the $\alpha$ clusters). The obtained distributions
were fitted by Gaussian functions to estimate their mean values
($\mu_{\gamma(\beta)}$, $\mu_{\alpha}$) and standard deviations
($\sigma_{\gamma(\beta)}$, $\sigma_{\alpha}$). After defining

$$
{\rm DP} = (\mu_{\gamma(\beta)} - \mu_{\alpha})/\sqrt{\sigma^2_{\gamma(\beta)}+\sigma^2_{\alpha}} \ ,
$$

\noindent as usually done for scintillating bolometers~\cite{Artusa:2014}, we obtain DP = 17(1) in an energy interval which includes $Q_{2\beta}$ of $^{116}$Cd. This high value for the DP, which can be even improved in underground conditions, is compatible with a full suppression of $\alpha$-induced background in the $0\nu2\beta$ decay ROI of $^{116}$Cd.

\begin{figure}[htb]
\centering
\includegraphics[width=0.48\textwidth]{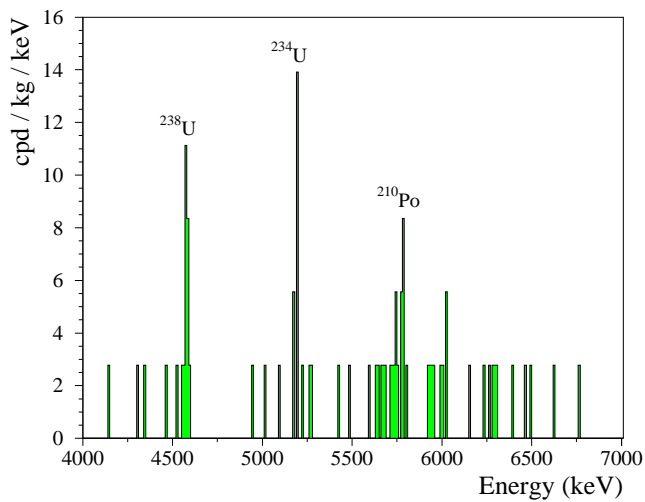}
\caption{Energy spectra of $\alpha$ events accumulated by the 34.5
g $^{116}$CdWO$_4$ bolometer over 250 h of data taking. The energy
scale corresponds to $\gamma$ energy calibration. The bin width is
10~keV. The $\sim 6-7$\% shift of the $\alpha$ peaks from the nominal
$Q_{\alpha}$ values is caused by a thermal quenching (see details
in Ref.~\cite{Arnaboldi:2010}).} \label{fig:alphas}
\end{figure}

The radioactive contamination of the $^{116}$CdWO$_4$ crystal was
estimated by using the energy spectrum of the $\alpha$ events,
presented in Fig.~\ref{fig:alphas}. The events were selected under
the condition that the associated LY be below 10 keV/MeV. The
peaks of $^{238}$U, $^{234}$U, and $^{210}$Po were identified in
the data. The $\alpha$ events outside the energy regions expected for U/Th
with their daughters can be explained by a surface pollution of the
$^{116}$CdWO$_4$ detector or/and of the surrounding construction
materials (which did not undergo an accurate cleaning process). Therefore, we have estimated the specific activities of the nuclides, while for other members of the U/Th chains only limits were obtained by using the procedure
recommended by Feldman and Cousins \cite{Feldman:1998}. The
estimations of the $^{116}$CdWO$_4$ crystal scintillator
radioactive contamination are presented in Table
\ref{tab:rad-cont}. Data on radioactive contamination of the
$^{116}$CdWO$_4$ crystal No.~1 described in Ref.~\cite{Poda:2014} are also reported.

\begin{table}[!htb]
\caption{Radioactive contamination of the $^{116}$CdWO$_4$ crystal
scintillator. Data on the radioactive contamination of the
$^{116}$CdWO$_4$ crystal No.~1 \cite{Poda:2014} are given for
comparison.} \footnotesize \centering

\begin{center}
\begin{tabular}{|c|c|c|c|}

\hline
 Chain      & Nuclide       & \multicolumn{2}{|l|}{Activity (mBq/kg)} \\
\cline{3-4}
 ~          & (sub-chain)   & This work & No.~1 \cite{Poda:2014} \\
\hline
 $^{232}$Th & $^{232}$Th    & $\leq0.13$        &   \\
 ~          & $^{228}$Th    & $\leq0.07$        & $0.031(3)$  \\
 $^{238}$U  & $^{238}$U     & $0.3(1)$          & $0.5(2)$ \\
 ~          & $^{234}$U     & $0.26(9)$         &   \\
 ~          & $^{230}$Th    & $\leq0.07$        &  \\
 ~          & $^{226}$Ra    & $\leq0.07$        & $\leq 0.005$ \\
 ~          & $^{210}$Po    & $0.23(8)$         & $0.6(2)$ \\
  $^{235}$U & $^{235}$U     & $\leq0.13$        &  \\
 \hline
\end{tabular}
\end{center}
\label{tab:rad-cont}
\end{table}

It should be noted that the $^{116}$CdWO$_4$ sample No.~1 was cut
from the same crystal boule, however our sample was closer to the
beginning of the boule. Therefore, the hint of a lower specific
activity of $^{238}$U and $^{210}$Po in the present sample (see Table \ref{tab:rad-cont}) can be explained by segregation of uranium and lead ($^{210}$Po being originated by $^{210}$Pb) in the CdWO$_4$
crystal growth process. As it was observed in Refs.
\cite{Barabash:2011,Danevich:2013,Poda:2013} the radioactive
contamination of the crystal boule by $^{228}$Th increases along
the boule from the growth cone to the bottom. Besides, the
contamination of the residuals after the crystal growth by
potassium, radium and thorium exceeds the boule contamination
significantly. These features indicate a strong segregation of the
radioactive impurities in the CdWO$_4$ crystal growing process.
Moreover, the radioactive contamination of sample No.~3 cut from
the $^{116}$CdWO$_4$ boule (here we again refer the reader to Fig. 1
in Ref.~\cite{Barabash:2016}) was significantly improved (in
particular, by one order of magnitude in thorium) after
recrystallization by the low-thermal-gradient Czochralski
method~\cite{Barabash:2016}. These results demonstrate encouraging
prospects for an enriched CdWO$_4$ crystal-scintillator production
with a radiopurity level satisfying the requirements of a
next-generation bolometric experiment.

\section{Conclusions}

A cadmium tungstate crystal scintillator with a mass of 34.5~g, enriched in $^{116}$Cd to 82\%, was tested over $\sim 250$~h at 18 mK as
a scintillating bolometer in an above-ground cryogenic laboratory.
The $^{116}$CdWO$_4$ detector exhibits a high energy resolution
(FWHM $\approx 2-7$~keV for $0.2-2.6$~MeV $\gamma$ quanta), and almost
complete discrimination between $\beta$($\gamma$) and $\alpha$
events (a discrimination power of $\sim 17$ was achieved in the
$2.6-7.0$~MeV region). These remarkable results were obtained in
spite of a significant pile-up effect related to the above-ground
location of the set-up.

We have found that the energy-to-voltage conversion and the time characteristics of the $^{116}$CdWO$_4$ signals are similar to those observed earlier with CdWO$_4$-based bolometers not produced from enriched material and sharing an akin detector design. The light yield observed in the present investigation is about twice higher (31 keV/MeV for $\gamma$ quanta) than that
given in the literature for CdWO$_4$ scintillating bolometers
thanks to the high optical quality of the enriched scintillator
and an efficient collection of the scintillation light in the detector
module.

The radioactive contamination of the $^{116}$CdWO$_4$ crystal by
$^{238}$U, $^{234}$U, and $^{210}$Po is estimated to be on the
level of $\sim0.3$ mBq/kg each, which is lower than that in the
$^{116}$CdWO$_4$ crystal samples cut from the same crystal boule
farther away from the growth cone (from which the studied sample was
obtained). This observation indicates a segregation of uranium and lead
in the CdWO$_4$ crystals growth process. For other $\alpha$
emitters belonging to the U/Th chains only limits on the level of
$0.07-0.13$~mBq/kg were obtained.

The present work demonstrates that $^{116}$CdWO$_4$ scintillating
bolometers represent one of the most promising technologies for
a next-generation bolometric experiment aiming at exploring the inverted
hierarchy region of the neutrino mass, as discussed in the
CUPID project.

\begin{acknowledgements}
The group from the Institute for Nuclear Research (Kyiv, Ukraine) was supported in part by the IDEATE International Associated Laboratory (LIA). The
researches were supported in part by the joint scientific project
``Development of Cd-based scintillating bolometers to search for
neutrinoless double-beta decay of $^{116}$Cd'' in the framework of
the PICS (Program of International Cooperation in Science) of CNRS
in years 2016-2018. This work was also supported by a public grant overseen by the French National Research Agency (ANR) as part of the ``Investissement d'Avenir'' program, through the IDI 2015 project funded by the IDEX Paris-Saclay, ANR-11-IDEX-0003-02. 
\end{acknowledgements}

\end{document}